\documentclass[prl,aps,twocolumn,showpacs,nobibnotes]{revtex4}
\usepackage{graphicx}
\usepackage{verbatim}

\begin{document}

\title{Power-law Temperature Dependent Hall Angle in the Normal State and its Correlation with Superconductivity in iron-pnictides}

\author{Y. J. Yan$^{1}$, A. F. Wang$^{1}$, X. G. Luo$^{1}$, Z. Sun$^{2}$$^\dag$, J. J. Ying$^{1}$, G. J. Ye$^{1}$, P. Chen$^{1}$, J. Q. Ma$^{1}$ and X. H. Chen$^{1}$}
\altaffiliation{E-mail of X.H.C: chenxh@ustc.edu.cn\\$^\dag$ E-mail
of Z.S: zsun@ustc.edu.cn} \affiliation{$^{1}$Hefei National
Laboratory for Physical Science at Microscale and Department of
Physics, University of Science and Technology of China, Hefei, Anhui
230026, People's Republic of China\\$^{2}$National Synchrotron
Radiation Laboratory, University of Science and Technology of China,
Hefei, Anhui 230029, People's Republic of China}

\begin{abstract}
We report Hall measurement of the normal state in K- and Co-doped
BaFe$_2$As$_2$, as well NaFe$_{1-x}$Co$_x$As. We found that a
power-law temperature dependence of Hall angle, cot$\theta_{\rm
H}$$\propto$ $T^\beta$, prevails in normal state with temperature
range well above the structural, spin-density-wave and
superconducting transitions for the all samples with various doping
levels. The power $\beta$ is nearly 4 for the parent compounds and
the heavily underdoped samples, while around 3 for the
superconducting samples. The $\beta$ suddenly changes from 4 to 3 at
a doping level that is close to the emergence of superconductivity.
It suggests that the $\beta$ of $\sim 3$ is clearly tied to the
superconductivity. Our data suggest that, similar to cuprates, there
exists a connection between the physics in the normal state and
superconductivity of iron-pnictides. These findings shed light on
the mechanism of high-temperature superconductivity.
\end{abstract}



\vskip 300 pt

\maketitle

As a second family of high-T$_c$ superconducting materials,
iron-pnictides have been frequently compared to high-T$_c$ cuprates
\cite{Hosono,Chen,JPaglione,GRStewart}. There are many similarities
between them, such as: an antiferromagentism  in the parent
compounds and the quasi-two-dimensional nature of superconducting
$CuO_2$ and FeAs layers. Superconductivity is realized by
suppressing the antiferromagnetic (AFM) ground state in both of
these superconductors. In cuprates, one of the central puzzles is
the unusual properties of the normal states, for instance,
pseudogap, linear-temperature dependent resistivity, $T^2$ behavior
of Hall angles, which can give clues to the underlying microscopic
interactions and the mechanism of superconductivity
\cite{Partrick,Chien}. Comparing the two high-T$_c$ families, we are
curious about whether the normal states in iron-pnictides can
provide some hints and help to uncover the high-T$_c$ physics. In
deed, unusual behavior, such as linear-temperature dependence of
magnetic susceptibility above the AFM transition
\cite{WangXF,WangXF1}, strong temperature-dependent Hall
coefficients \cite{Rullier,Fang,Kasahara,Pallecchi,Aswartham}, has
been observed in iron-pnictides. Although the underlying physics is
still under debate, these properties are closely related to the
multiband character of iron-pnictides that is a crucial key to the
understanding of the superconductivity in these materials.

While de Haas-van Alphen and angle-resolved photoemission
spectroscopy (ARPES) can precisely determine the Fermi surface
topology and the band structure, the electronic transport
measurements are more sensitive to the subtle and complicated
interactions in the multiband system. In particular, transport
measurements can determinately uncover anomalous behavior in the
normal states. Here, we will show that resistivity and Hall
measurements hint a connection between the high-temperature normal
state and the low-temperature superconducting state. We have
investigated a series of single crystals of
Ba$_{1-x}$K$_x$Fe$_2$As$_2$ and Ba(Fe$_{1-x}$Co$_x$)$_2$As$_2$, as
well NaFe$_{1-x}$Co$_x$As. The cotangent of Hall angle,
cot$\theta_{\rm H}$, from resistivity and Hall measurements was
observed to vary with a simple but systematic trend. The
cot$\theta_{\rm H}$ shows a fashion of $T^\beta$ in the paramagnetic
state well above the structural, SDW and superconducting
transitions. The magnitude of $\beta$ is $\sim$ 4 in the heavily
underdoped regime near the parent compound, while it drops to $\sim$
3 when the superconducting ground state emerges. With further
doping, $\beta$ remains $\sim$ 3 in a small doping range and then
decreases gradually with the vanishing of superconducting ground
state. Together with the similar behavior observed in
NaFe$_{1-x}$Co$_x$As, our data show a consistent behavior in the
high-temperature behavior in the normal state, which bears a
connection with the emergence of superconducting ground state.

\begin{figure}[t]
\centering
\includegraphics[width=0.5 \textwidth]{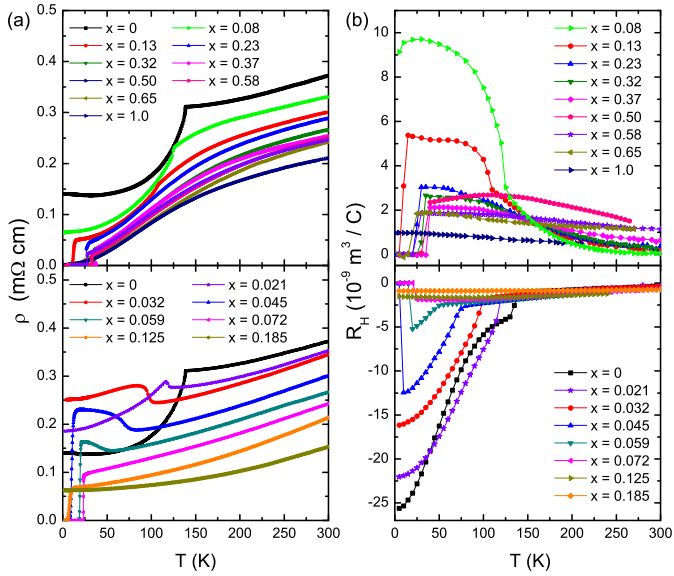}
\caption{(Color online) (a): Temperature dependence of resistivity
for Ba$_{1-x}$K$_x$Fe$_2$As$_2$ (top) and
Ba(Fe$_{1-x}$Co$_x$)$_2$As$_2$ (bottom) single crystals,
respectively; (b): Temperature dependence of Hall coefficient for
Ba$_{1-x}$K$_x$Fe$_2$As$_2$ (top) and Ba(Fe$_{1-x}$Co$_x$)$_2$As$_2$
(bottom) single crystals, respectively.} \label{fig1}
\end{figure}

\begin{figure}[t]
\centering
\includegraphics[width=0.5 \textwidth]{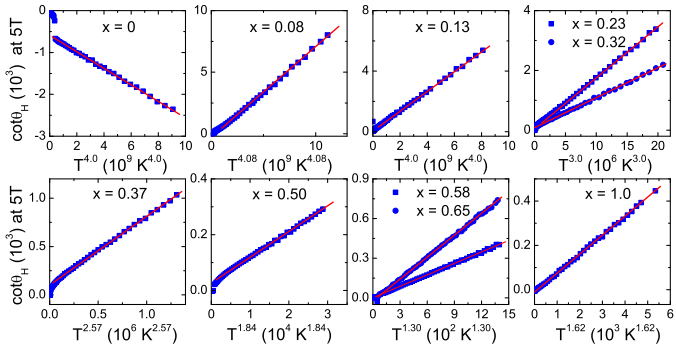}
\caption{(Color online) cot$\theta_{\rm H}$ $vs.$ $T^{\beta}$ for
the single crystals of Ba$_{1-x}$K$_x$Fe$_2$As$_2$ system with
different x. The magnitude of $\beta$ varies from sample to sample.
The solid red lines show the $T^{\beta}$ dependence of
cot$\theta_{\rm H}$.} \label{fig2}
\end{figure}

\begin{figure}[t]
\centering
\includegraphics[width=0.5 \textwidth]{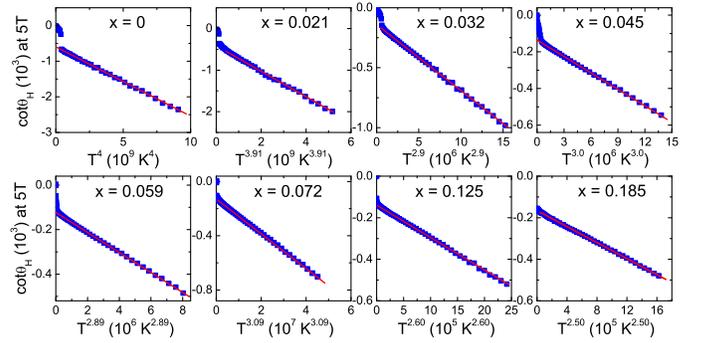}
\caption{(Color online) cot$\theta_{\rm H}$ $vs.$ $T^{\beta}$ for
Ba(Fe$_{1-x}$Co$_x$)$_2$As$_2$ single crystals with various doping.
The magnitude of $\beta$ varies from sample to sample. The solid red
lines show the $T^{\beta}$ dependence of cot$\theta_{\rm H}$.}
\label{fig3}
\end{figure}

\begin{figure}[t]
\centering
\includegraphics[width=0.5 \textwidth]{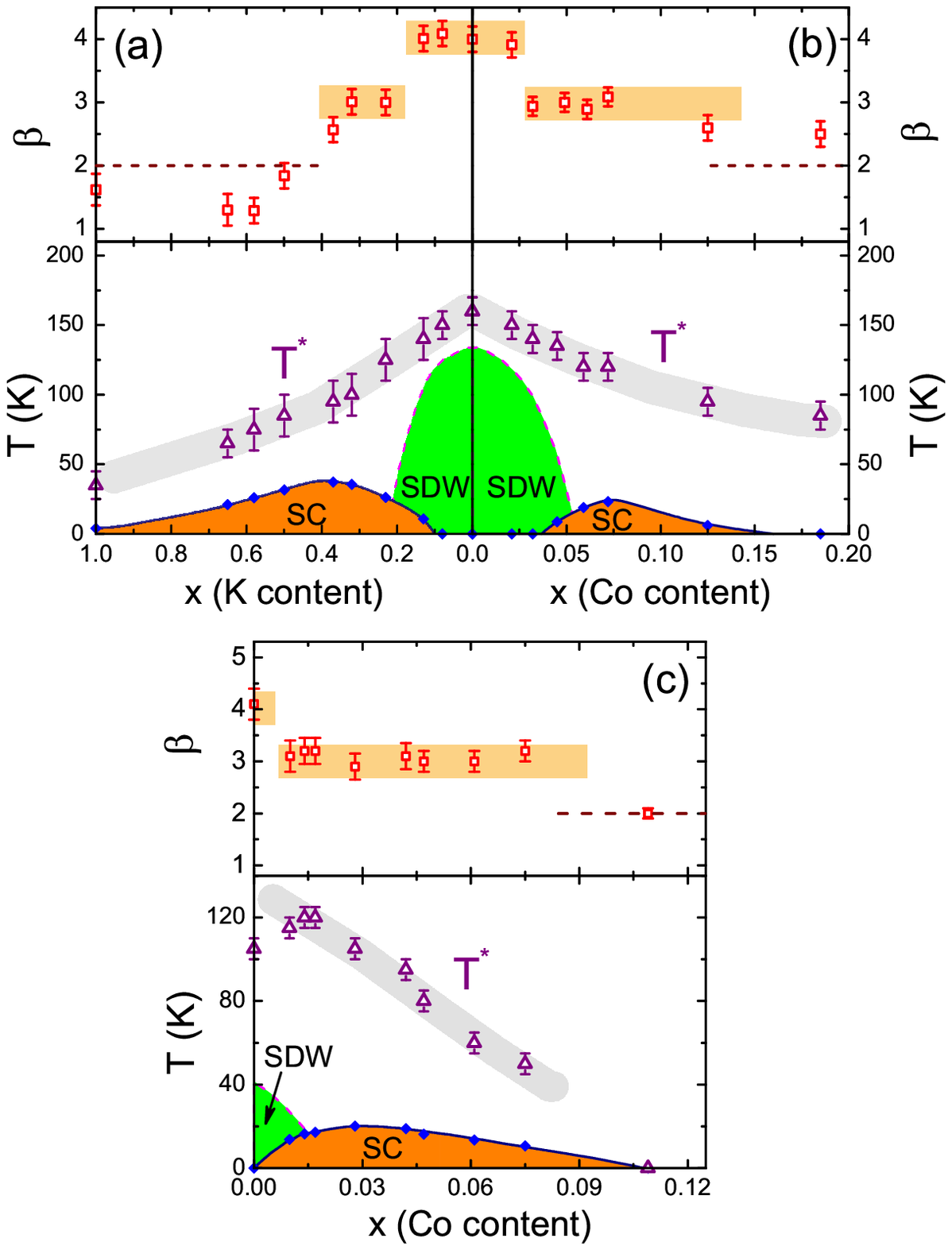}
\caption{(Color online) The doping dependence of the power-law
exponent $\beta$ and phase diagrams for \textbf{(a):}
Ba$_{1-x}$K$_x$Fe$_2$As$_2$, \textbf{(b):}
Ba(Fe$_{1-x}$Co$_x$)$_2$As$_2$ and \textbf{(c):}
NaFe$_{1-x}$Co$_x$As, respectively. The data of NaFe$_{1-x}$Co$_x$As
are taken from Ref.17. The $T^{\ast}$ is the characteristic
temperature at which the Hall angles deviate from the
high-temperature $T^{\beta}$ behavior.} \label{fig4}
\end{figure}

Figure 1(a) shows the temperature dependence of resistivity for
single crystalline Ba$_{1-x}$K$_x$Fe$_2$As$_2$ and
Ba(Fe$_{1-x}$Co$_x$)$_2$As$_2$ with various doping, with a marked
asymmetric changes of transport properties induced by electron and
hole doping. Our data are similar to previous reports of resistivity
in doped BaFe$_2$As$_2$ crystals \cite{Fisher,ChenH}. It is evident
that the high-temperature resistivity exhibits distinct curvatures
for electron and hole dopings. In the hole-doped region (Fig. 1(a),
top panel), the high-temperature curvature is downwards, while it is
upwards in the electron-doped region (Fig. 1(a), bottom panel). It
is interesting that the high-temperature resistivity of the parent
compound bears a similarity to that of electron-doped compounds,
which suggests that at high temperatures the parent compound can be
considered as an electron doped compound. This argument is further
reinforced by the results of Hall measurements, which shows that at
high temperatures the electron carriers dominate the transport
properties in the parent compound (Figure 1(b)).

On the other hand, the transport measurements show complex behavior
at low temperatures below the structural and SDW transitions. The
resistivity shows different temperature-dependence in the Co and K
doping regimes. In fact, it has been found that SDW, electronic
nematicity and orbital ordering etc. take place below the
structural/SDW transitions \cite{Fisher1}, which can significantly
reconstruct the band structure and give rise to novel properties in
the electronic systems \cite{Fisher1}. In this paper, we focus on
the data at high temperatures well above the structural and SDW
transitions. We stress that, without the driving forces from those
ordering tendency, the high-temperature transport data are able to
provide information on the fundamental changes in the prime band
structure, which can supply valuable information on the physics of
iron-pnictides.

Figure 1(b) shows the Hall coefficient $R_{\rm H}$ for both
Ba$_{1-x}$K$_x$Fe$_2$As$_2$ (top) and Ba(Fe$_{1-x}$Co$_x$)$_2$As$_2$
(bottom) crystals with various doping. In a typical compensated
semimetal, where the densities of electron and hole carriers are
roughly equal to each other, a vanishing Hall coefficient $R_{\rm
H}$ is usually expected. However, similar to earlier reports by
other groups \cite{Rullier,Fang,Kasahara,Pallecchi,Aswartham}, the
Hall coefficient  $R_{\rm H}$ shows a strong temperature dependence.
In particular, the transport properties seem to be dominated by a
single type of carriers in these multiband systems, and a remarkable
electron or hole character of transport properties can be induced by
a slight electron or hole doping. It is evident that the $R_{\rm H}$
of Ba(Fe$_{1-x}$Co$_x$)$_2$As$_2$ is always negative, suggesting the
dominance of electron carriers. In Ba$_{1-x}$K$_x$Fe$_2$As$_2$, the
$R_{\rm H}$ turns to be positive with slight hole doping ($x \geq
0.08$), showing that the hole carriers govern the transport
properties. The significant rise in the magnitude of $R_{\rm H}$ is
associated with the structural/SDW transitions, which is not the
focus of the our study here. How to explain the unconventional Hall
properties is still under debate. In this paper, we focus on the
high temperature region well above the structural, SDW and
superconducting transitions.

Despite of its complicated properties, we found that the resistivity
and Hall data reveal a intrinsic but simple behavior. Using the
resistivity and Hall data displayed in Figs. 1(a) and (b), we can
calculate the cotangent of Hall angles, cot$\theta_{\rm
H}$=$\rho/\rho_{xy}$, for both Ba$_{1-x}$K$_x$Fe$_2$As$_2$ and
Ba(Fe$_{1-x}$Co$_x$)$_2$As$_2$ crystals. The Hall angle reveals a
power-law temperature dependence, cot$\theta_{\rm H}$ = $A$ + $B
T^\beta$, in the temperature range well above SDW, structural and
superconducting transitions. Figures 2 and 3 show the plots of
cot$\theta_{\rm H}$ vs. $T^{\beta}$ for both
Ba$_{1-x}$K$_x$Fe$_2$As$_2$ and Ba(Fe$_{1-x}$Co$_x$)$_2$As$_2$,
respectively. A $T$-power law dependence holds for all the crystals
at high temperatures. In Figure 2, the cot$\theta_{\rm H}$ of the
parent compound BaFe$_2$As$_2$ decreases with increasing
temperature, which is similar to the behavior of electron-doped
compounds as shown in Fig. 3.  Together with the resistivity data,
this behavior suggests that BaFe$_2$As$_2$ can be considered as an
electron doped compound in the high-temperature normal state. In
contrast, the cot$\theta_{\rm H}$ in hole-doped samples for $x \geq
0.08$ increases with temperature. Our data indicate that the
temperature dependence of the cot$\theta_{\rm H}$ can provide
information on whether electron or hole carriers predominate the
transport properties. Moreover, one may notice that the magnitude of
the power $\beta$ varies from sample to sample. Moreover, such a
$T^{\beta}$ behavior of Hall angle in the normal state has also been
found in Co-doped NaFeAs family, as reported by us in Ref. 17.

For a comparison, the $T^{\beta}$ behavior of Hall angle has been
observed in cuprate superconductors, which is considered to be a
peculiar properties of the unusual normal state, though the
explanation for this behavior is controversial. In most cuprates,
$T^2$ dependence of the cot$\theta_{\rm H}$ holds for a wide doping
range in the normal state above the pseudogap-opening temperature
\cite{Chien,Ando}. In contrast, our data show that the power $\beta$
varies with doping in iron-pnictides. Together with the phase
diagrams, the powers $\beta$ for the K- and Co-doped BaFe$_2$As$_2$
and Co-doped NaFeAs are summarized in Fig. 4. Moreover, we have
marked in the phase diagrams the $T^{\ast}$ temperatures, below
which the Hall angles deviate from the $T^{\beta}$ behavior. It is
very interesting that the evolutions of $T^{\ast}$ with doping show
a highly consistent behavior in both of the doped BaFe$_2$As$_2$ and
NaFeAs families, and further studies are required to show what
happens at these crossover temperatures. Systematic measurements by
ARPES have shown that the carrier doping induced by K and Co dopants
leads to a rigid-band-like change of the valence band structure
\cite{LiuC,Neupane,Ideta}. Therefore, Fig. 4 displays a systematic
change of electronic properties due to the variation of electron and
hole carriers in BaFe$_2$As$_2$ and NaFeAs. It is worth noting that
the power-law behavior occurs in high temperature region above the
structural, SDW and superconducting transitions, which reveals the
fundamental electronic properties of the iron-pnictides without the
complexity due to the low-temperature electronic reconstructions.

In Fig. 4, it is remarkable that the variation of the power $\beta$
behaves in a highly similar fashion in BaFe$_2$As$_2$ and NaFeAs
families. In the parent compounds of BaFe$_2$As$_2$ and NaFeAs,
$\beta$ $\sim$ 4, and the magnitude of $\beta$ persists until the
emergence of superconductivity. In the doping range where the
superconducting ground state prevails, $\beta$ drops to $\sim$ 3.
Then, $\beta$$\sim$3 holds for a wide range of superconducting
compounds. With further doping, superconductivity fades away,
$\beta$ decreases to a smaller value. With the emphasis on the
consistent variation of $\beta$ in these materials, we also notice
some exceptions. In the parent compound of NaFeAs family, there is
no bulk superconductivity, we therefore consider it a
non-superconducting compound. In addition, the significant decrease
of $\beta$ in the overdoped Ba$_{1-x}$K$_x$Fe$_2$As$_2$ is likely
coincident with the fact that the isotropic gap structure gradually
changes to a nodal one with K doping\cite{DingHX,LiSY}, though it is
unclear how such a variation can change the magnitude of $\beta$.

As already shown in Fig. 1, the electronic transport in electron and
hole doped region of BaFe$_2$As$_2$ is quite different. Moreover,
many properties of NaFeAs family are distinct from those of
BaFe$_2$As$_2$ compounds. However, the similar doping-dependent
variation of $\beta$ in these materials suggests that there is a
connection between the high-temperature transport data and the
low-temperature electronic ground states. At high temperatures, the
electronic band structure and underlying interactions are relatively
simple, without the complexity due to the electronic reconstruction
or ordering tendency at low temperatures. With the electron and hole
doping, ARPES data have shown that the band structure changes in a
rigid-band-like fashion, without significant variation of local
correlations \cite{LiuC,Neupane,Ideta}. Thus, it is reasonable to
believe that, at high temperatures, the fundamental change in the
electronic system is the shifting of Fermi energy. Generally, one
may expect $\beta$ evolves gradually with carrier doping, since the
rigid-band-like change in the band structure is smooth
\cite{LiuC,Neupane,Ideta}. However, the reduction of $\beta$ from 4
to 3 seems to be a sudden drop, which suggests a particular change
of the band structure by either hole or electron doping. It is worth
noting that, the electronic properties of electron and hole pockets
in iron-pnictides are remarkably different, while our data shows
that the Hall angles in the electron- and hole-doped BaFe$_2$As$_2$
behave in a similar fashion. This counterintuitive finding suggests
that the normal-state electronic structure and its interactions with
various degrees of freedom contain important messages about
iron-pnictide physics. Unfortunately, there is no high-temperature
ARPES data available to unambiguously show the critical change in
the band structure \cite{LiuC,Neupane,Ideta}. We can only infer such
a change by the variation of $\beta$. The close connection between
this reduction of $\beta$ and the emergence of superconducting
ground state suggests that the change of the band structure by
doping favors the superconductivity in iron-pnictides. Moreover,
Fig. 3 shows that $\beta$$\sim$3 is associated with the
superconducting ground state, though $\beta$ decreases continuously
with K doping in overdoped Ba$_{1-x}$K$_x$Fe$_2$As$_2$, which is
probably associated with the variation of the gap structure from
isotropic to nodal one. In particular in Co-doped BaFe$_2$As$_2$ and
NaFeAs, $\beta$$\sim$3 prevails in the whole superconducting
regimes. All these properties point out the high-temperature normal
state possesses substantial clues on the superconductivity in
iron-pnictides, though it is unclear how the change of the band
structure boosts superconductivity in cooperation with spin or
orbital degrees of freedom.

Even though the Hall measurements of BaFe$_2$As$_2$ and NaFeAs
families show unconventional properties that has not been
unambiguously explained, the power-law temperature dependence of
Hall angles reveal that  $\beta$ $\sim$ 3 is crucial for
superconductivity, which could give a clue on the connection between
superconductivity and the complex interactions in iron-pnictides.
The electron doped and hole doped iron-pnictide superconductors
exhibit distinct Fermi surface topology\cite{LiuC,Neupane,Ideta},
thus one would expect the different transport properties.
Surprisingly, the power-law temperature dependence of Hall angles
$\beta$ $\sim$ 3 in the normal state above a characteristic
temperature ($T^*$) is universal for the superconducting samples.
The intrinsic mechanism of this unexpected power-law behavior in the
normal state is still unknown, and further works need to do to
unveil its origin. Moreover, superconductivity has been found in Ru
and P doped BaFe$_2$As$_2$, in which Ru and P change local
interactions in the electronic system. It would be interesting to
investigate whether $\beta$ $\sim$ 3 is a universal behavior and
holds for these materials. These studies, together with our results,
will shed light on the roles of charge itinerancy and local
interactions in iron-pnictides.

In summary, the Hall angle of Ba$_{1-x}$K$_x$Fe$_2$As$_2$,
Ba(Fe$_{1-x}$Co$_x$)$_2$As$_2$ shows a power-law behavior,
cot$\theta_{\rm H}$$\propto$ $T^\beta$, at high temperatures well
above the structural, SDW and superconducting transitions. $\beta$
$\sim$ 4 was observed for the parent compound and the heavily
underdoped samples on both the electron and hole doped sides. With
increasing doping, as the superconductivity occurs, the $\beta$
suddenly decreases to $\sim$ 3. The close connection between the
change of $\beta$ magnitude and the emergence of superconductivity
suggest that some important electronic property at high temperatures
is crucial for the understanding of the superconductivity in
iron-pnictides.

\textbf{Acknowledgements:}

This work is supported by the National Natural Science Foundation of
China (Grants No. 11190021, 11174266, 51021091), the "Strategic
Priority Research Program (B)" of the Chinese Academy of Sciences
(Grant No. XDB04040100), the National Basic Research Program of
China (973 Program, Grants No. 2012CB922002 and No. 2011CBA00101),
and the Chinese Academy of Sciences.

\end{document}